# Double-Strand Break Clustering: An Economical and Effective Strategy for DNA Repair

Junyi Chen[1], Wenzong Ma, Yuqi Ma[1], Gen Yang[1,2]*

[1] State Key Laboratory of Nuclear Physics and Technology, School of Physics, Peking University, Beijing 100871, P. R. China
[2] Oncology Discipline Group, the Second Affiliated Hospital of Wenzhou Medical University, Wenzhou 325003, P. R. China

*Corresponding author
Correspondence to: Gen Yang (gen.yang@pku.edu.cn)

## Abstract

In mammalian cells, repair centers for DNA double-strand breaks (DSBs) have been identified. However, previous researches predominantly rely on methods that induce specific DSBs by cutting particular DNA sequences. The clustering and its spatiotemporal properties of non-specifically DSBs, especially those induced by environmental stresses such as irradiation, remains unclear. In this study, we used Dragonfly microscopy to induce high-precision damage in cells and discovered that DSB clustering during the early stages of DNA damage response (DDR) and repair, but not during the repair plateau phase. Early in DDR, DSB clustered into existing 53BP1 foci. The DSB clustering at different stages has different implications for DNA repair. By controlling the distance between adjacent damage points, we found that the probability of DSB clustering remains constant at distances of 0.8 - 1.4 μm, while clustering does not occur beyond 1.4 μm. Within the 0.8 μm range, the probability of clustering significantly increases due to the phase separation effect of 53BP1. Using a Monte Carlo approach, we developed a dynamic model of 53BP1 foci formation, fission, and fusion. This model accurately predicts experimental outcomes and further demonstrates the temporal and spatial influences on DSB clustering. These results showed that, similarly to specifically induced DSBs, non-specifically induced DSBs can also cluster. The extent of DSB clustering is influenced by both temporal and spatial factors, which provide new insights into the dynamics of DSB clustering and the role of 53BP1 in DNA repair processes. Such findings could enhance our understanding of DNA damage responses and help us improve DNA repair therapies in disease.
**Keywords**: DSB cluster, DNA repair, Computational biophysics

# Introduction

Among the various types of DNA damage, double-strand breaks (DSBs) are the most severe, since unrepaired DSBs can lead to genomic instability, malignant tumors, and even cell death. Therefore, investigating the specific repair process of DSBs is crucial. DSBs exhibit a certain degree of mobility, with ongoing debate regarding whether their movement is directional or random[1-4]. Increasing evidence suggests the existence of DSB repair centers, where DSBs cluster on a large scale[5-10]. Given that many DNA repair-related proteins are multivalent[11-13], clustering allows cells to repair more DSBs with fewer proteins, which offers a more economical and efficient strategy for cells, aligning with the principles of biological evolution. Additionally, while DSB clustering can enhance the accuracy of repair, it also increases the risk of chromosomal translocations[6,7]. Current evidence for DSB clustering mainly comes from specific systems where particular DNA sequences are cleaved to induce specific DSBs[7,9,10], and there still lacks evidence of DSB clustering in non-specific systems, such as random damage caused by radiation.

The DNA damage response (DDR) pathway consists of sensors, transducers, and effectors[14-16]. Upon detecting DNA damage, cells recruit 53BP1 via ATM[17], creating DNA repair compartments for DSBs[18] and simultaneously recruiting downstream repair proteins. The DDR process exhibits different characteristics as it progresses over time. DSB clustering is associated with the ATM and MRN complexes that detect DSBs[7,10], underscoring its role in DDR. Currently, the most effective method to studying DSB clustering is the induction of site-specific DNA damage with endonucleases. Since the timing of enzyme expression and their entry into the nucleus vary, this significantly affects the induction time and repair kinetics of DSBs[19]. Therefore, it is challenging to determine at which stage of DDR the DSB clustering occurs, whether the degree of clustering changes over time, and whether the clustering at different stages has different effects on repair.

53BP1 is a pivotal signaling protein in the DDR pathway[17,20], which exhibits the property of liquid-liquid phase separation[12,18]. Due to its surface tension, there is a size limit for 53BP1. During DNA repair, the fusion of 53BP1 foci is a common observation, effectively demonstrating the clustering of DSBs[8]. It is conceivable that adjacent 53BP1 are more likely to cluster and merge because of surface tension[10], though this correlation remains unverified. Investigating the relationship between DSB clustering and spatial distance is essential to determine at which scale the surface tension of 53BP1 significantly influences clustering. Laser microirradiation has been widely applied for inducing DNA damage due to its high precision and capability for real-time observation[21-23]. This technique would be particularly useful for examining the dynamics and spatial characteristics of DSB clustering.

The Monte Carlo method serves as a robust tool for simulating the dynamic processes of DSB[24,25], offering profound insights into the underlying mechanisms behind the phenomena. The Monte Carlo method can provide data that may be challenging or impossible to obtain through experiments, thereby guiding experimental approaches. Its potential to explore the dynamics of DSB clustering is particularly promising. In our previous work, we established a single chromatin model that effectively simulates DNA damage post-irradiation[26,27]. Additionally, a regional model can be constructed to simulate the formation of foci[28,29]. However, current approaches to modeling the subsequent repair dynamics of these foci are limited to mathematical fitting techniques[5,30], and no simulation models have been established.

In this study, we employed the Micropoint for highly precise laser microirradiation on cells and revealed that non-specifically induced DSBs cluster. The degree of clustering varies over time and across different repair processes. Clustering during the early stages of DDR and DSB repair holds different implications for the repair process. Furthermore, by measuring changes in DSB clustering intensity at different distances, we identified at least two influencing factors, one of which is the phase separation of 53BP1 that promotes clustering at shorter distances. To delve deeper into the spatio-temporal specificity of DSB clustering, we also developed a dynamic model based on Monte Carlo calculation. This model, which details the formation, fission, and fusion of 53BP1 foci, effectively predicts the repair dynamics of radiation-induced foci, offering valuable guidance for future experimental research.

## Results

### DSBs clustered into existing 53BP1 foci

To explore whether non-specific double-strand breaks cluster, we utilized the Dragonfly microscope equipped with a high-energy UV laser, Micropoint, to precisely induce damage in cells. We utilized the HT1080 cells transiently transfected with 53BP1-ptdTomato to induce DSBs at specific locations. The resolution of the laser-induced damage closely matches the imaging resolution of the microscope. We first induced damage within the cells in a vertical 1×5 dot matrix, spacing the dots 2 μm apart vertically to ensure that the DSBs did not interact with each other. After 5 minutes, each dot position formed 53BP1 foci (Fig. 1a). We then induced damage in another identical vertical 1×5 dot matrix at 0.8μm on the right side of the original dot matrix. Theoretically, if there is no clustering of DSBs, a new column of foci should emerge on the right. However, we observed no new column of foci forming on the right. Instead, the fluorescence intensity of the original 53BP1 foci significantly increased (Fig. 1b), and there was no noticeable migration of the existing 53BP1 foci to the right. Interestingly, altering the spacing between the two dot matrices to 4μm resulted in a new column of foci on the right.

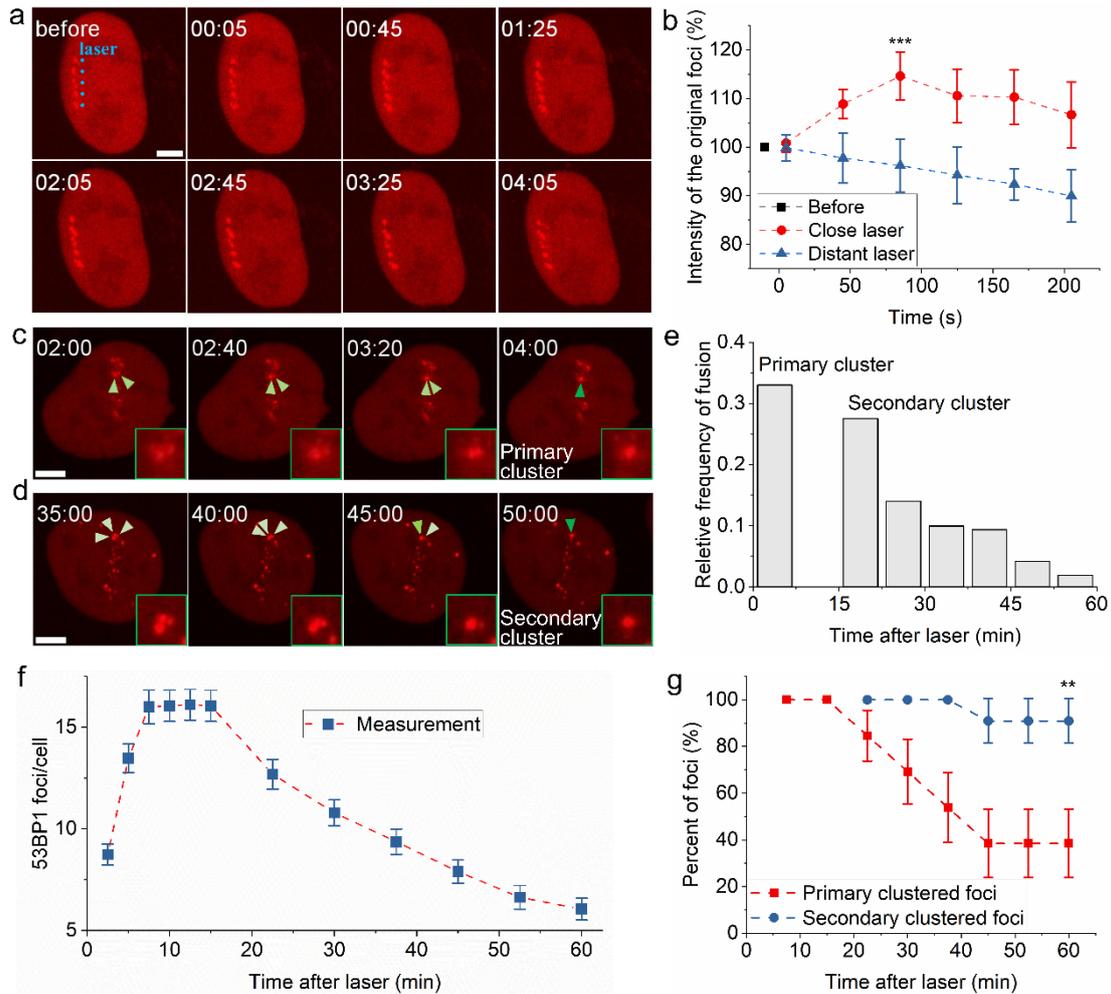

**Figure 1. Temporal specificity of DSB clustering.**
**a**, Representative time series image of a column of 1×5 damage points was induced 0.8 μm to the right of a column of 53BP1 foci; sacle bar, 5 μm. **b**, Fluorescence intensity curve of the original foci, data shown as mean and s.e.m (black, before another laser; red, after another close laser, n=8; blue, after another distant laser, n=11; *** $p<0.001$, independent samples t-test). **c**, Representative image of the fusion of primary clustered foci. Scale bar, 5 μm. **d**, Representative image of the fusion of secondary clustered foci. Scale bar, 5 μm. **e**, Distribution of fusion frequency over time (n=78); time scale, 7.5 min. **f**, Repair curve of foci, data shown as mean and s.e.m (n=19). **g**, Repair curves for primary and secondary clustered foci, data shown as mean and s.e.m (red, primary clustered foci, n=13; blue, secondary clustered foci in 15–22.5 min, n=11; ** $p<0.01$, independent samples t-test).

## DSB clustering varies with time

To further investigate at which stage of DNA repair the DSB clustering occurs, we utilized laser microirradiation on cells and monitored the dynamics of 53BP1 foci. We observed a rapid increase in the count of 53BP1 foci within 0-7.5 minutes post-irradiation (Fig. 1f), along with the occurrence of foci fusion (Fig. 1c). From 7.5 to 15 minutes, the count of foci remained relatively stable, with no further fusion or significant clustering observed. After 15 minutes, the foci began to repair, accompanied by fusion. As the number of foci decreased, the frequency of fusion also diminished (Fig. 1d, e). This suggests that DSB clustering may vary with the progression of the repair process. We refer to the clustering occurring at the early stage of DDR as primary clustering, and that occurring during the DSB repair phase as secondary clustering. Moreover, we noted that the primary clustered foci repair faster than the secondary clustered foci (Fig. 1g), and the repair rate of primary clustered foci is similar to that of non-clustered foci. This indicates distinct implications of primary and secondary clustering for DSB repair.

To further explore the temporal changes in the fusion of 53BP1 foci, HT1080 cells stably transfected with 53BP1-GFP were irradiated with $^{60}$Co. We again observed the fusion phenomena in 53BP1 foci (Supplementary Fig. 1a). Between 15 and 30 minutes post-irradiation, the count of 53BP1 foci remained essentially unchanged (Supplementary Fig. 1d). During this, few foci disappeared or newly appeared (Supplementary Fig. 1e), suggesting minimal DSB repair during this period. Additionally, the positions of foci were highly stable, exhibiting neither fusion nor obvious clustering (Supplementary Fig. 1b, c). After 30 minutes, the foci began to be repaired and fusion occured (Supplementary Fig. 1b, d).

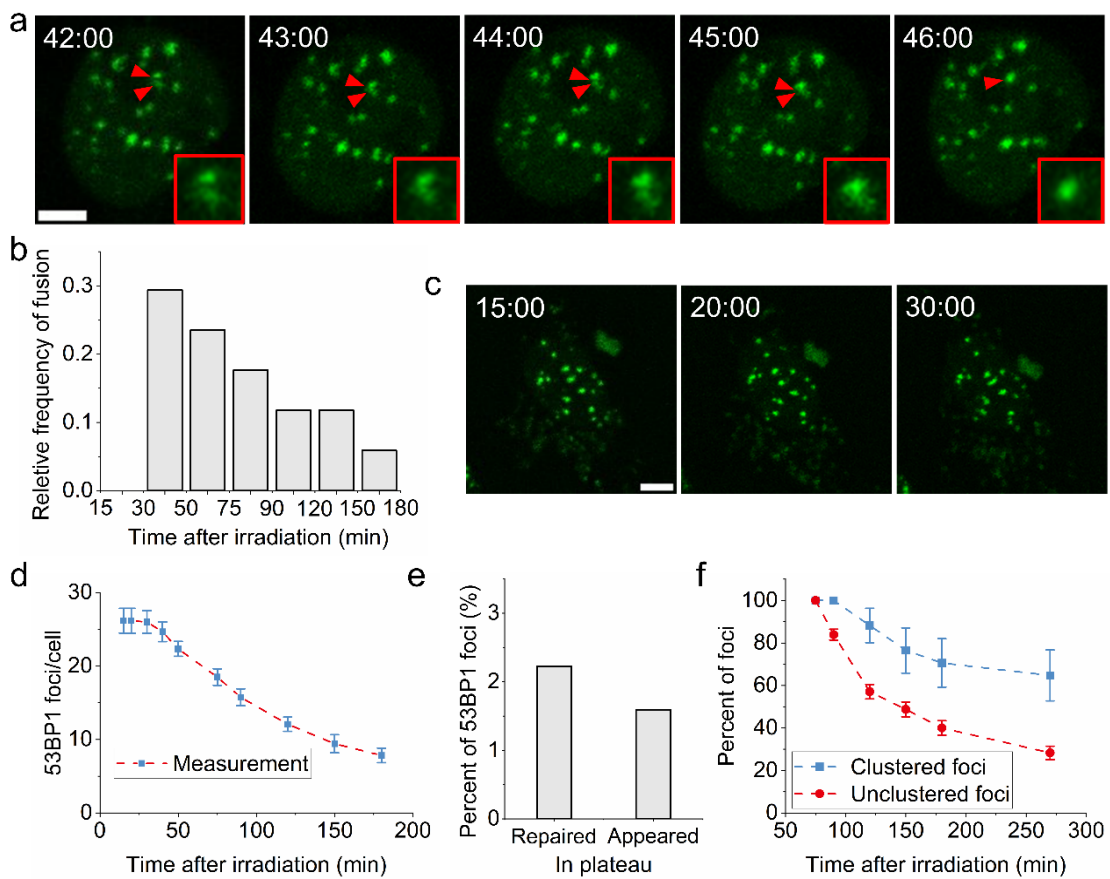

**Supplementary Figure 1. Temporal specificity of DSB clustering under $^{60}$Co Irradiation.**
**a**, Representative image of 53BP1 foci fusion, scale bar, 5 μm. **b**, Relative frequency distribution of foci fusion over time (n=38). **c**, Representative image of foci at rest during the plateau phase, scale bar, 5 μm. **d**, Repair curve of 53BP1 foci (n=12). **e**, During the plateau phase between 15-30 minutes post-irradiation, the number of foci remains constant, showing the ratio of repaired and newly appeared foci (n=314). **f**, Percentage repair curve of foci, data shown as mean and s.e.m. (blue, foci clustered 75 minutes prior, n=17; red, unclustered foci, n=205).

To determine whether DSBs cluster before 15 minutes, we irradiated the cells by stages. After delivering 1 Gy of irradiation and waiting 15 minutes, a second dose of 1 Gy was delivered, followed by another 15-minute to assess 53BP1 foci (Supplementary Fig. 2b). Given the minimal repair of DSBs between 15 and 30 minutes, the total number of DSBs generated by 1 Gy + 1 Gy irradiation was comparable to that from a single 2 Gy dose. This suggested clustering of DSBs generated by the second irradiation into the 53BP1 foci generated by the first. To exclude the possibility of 53BP1 depletion, cells were irradiated with doses ranging from 1 to 4 Gy (Supplementary Fig. 2a). At 2 Gy, 53BP1 levels were not depleted (Supplementary Fig. 2c, d), ruling out 53BP1 depletion as a cause for the observed effects under the 1 Gy + 1 Gy conditions. Collectively, these data indicate that DSBs cluster during the early DDR and DSB repair phase, whereas no clustering occurs during the plateau phase.

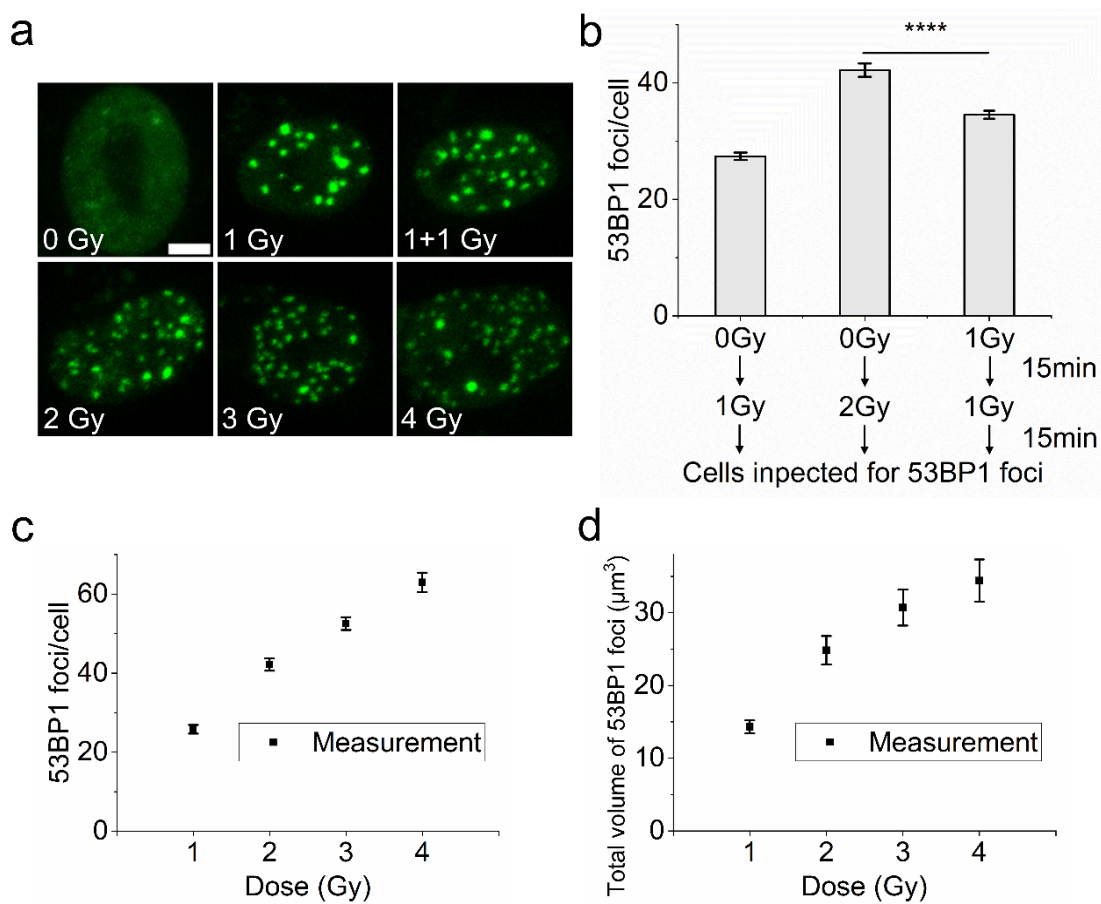

**Supplementary Figure 2. DSBs clustered into pre-existing 53BP1 foci during the early phase of DDR.**

**a**, Representative images of 53BP1 foci induced by $^{60}$Co radiation at 1-4 Gy doses and fractionated radiation of 1+1 Gy, scale bar, 5 μm. **b**, Number of 53BP1 foci under fractionated irradiation, data shown as mean and s.e.m (n≥261; **** p<0.0001, independent samples t-test). **c**, Increase in the number of 53BP1 foci with increasing doses, data shown as mean and s.e.m (n≥58). **d**, Total volume of 53BP1 foci in the entire nucleus, data shown as mean and s.e.m (n≥58).

## Spatial physical properties of DSBs clustering

To further explore the spatial characteristics of DSB clustering, we used Micropoint to induce damage in a vertical 2×5 dot matrix in cells (Fig. 2a, b). Since the vertical distance of 2 μm between damage points did not lead to interaction, we adjusted the horizontal intervals to investigate the phenomenon. Setting the 53BP1 foci observed at 10 min post-irradiation as standard. When the horizontal interval is 1.4-2 μm, all damage points form 53BP1 foci respectively (Fig. 2a). This suggests that in the absence of DSB clustering, foci will form at each damage point, with no fusion occuring. Reducing the horizontal intervals led to the fusion of foci (Fig. 1c), and at even closer intervals there was a probability of observing only one focus at two damage points (Fig. 2b). Thus, when reducing the transverse interval, the probability of observing only one focus can reflect the intensity of DSB clustering to a certain extent. When the horizontal interval was between 0.8 and 1.4 μm, the probability remained nearly constant at around 40%. However, when the horizontal interval is less than 0.8 μm, the probability will increase as the interval is reduced (Fig. 2c), indicating that the intensity of DSB clustering increases with decreasing distance.

To investigate the reasons for increased clustering intensity within a 0.8 μm interval, we introduced 1,6-hexanediol (Hex) into the cells. 1,6-Hex interferes with hydrophobic interactions among proteins, thereby constraining the 53BP1 droplets[12]. We treated HT1080 cells with 4 μM etoposide for 30 minutes to induce 53BP1-GFP. After adding 0.2% 1,6-Hex, there was a reduction in the fluorescence intensity of 53BP1 foci (Fig. 2d, e). Furthermore, when using Micropoint to induce a 2×5 dot matrix of damage, the probability of observing only one focus on both sides significantly reduced in the case of 0.6 μm interval, while for the intervals of 0.8 and 1.1 μm there was no significant change (Fig. 2f). It is noteworthy that the interval of 0.6 μm approximates twice the radius of 53BP1 foci.

To explore the impact of DNA mobility on DSB clustering from a physical perspective, we treated cells with KAT8-IN-1. KAT8-IN-1 inhibits KAT8, thereby reducing the acetylation of histone H4K16[31], leading to more compact DNA double strands and decreased mobility. According to current literature, there is no direct link between KAT8 and DNA repair. After introducing KAT8-IN-1 and inducing a 2×5 dot matrix of damage with a 0.8 μm horizontal interval, we observed more cases where foci were produced on both sides (Fig. 2g, h). This suggests that lower mobility of DNA strands leads to decreased intensity of DSB clustering.

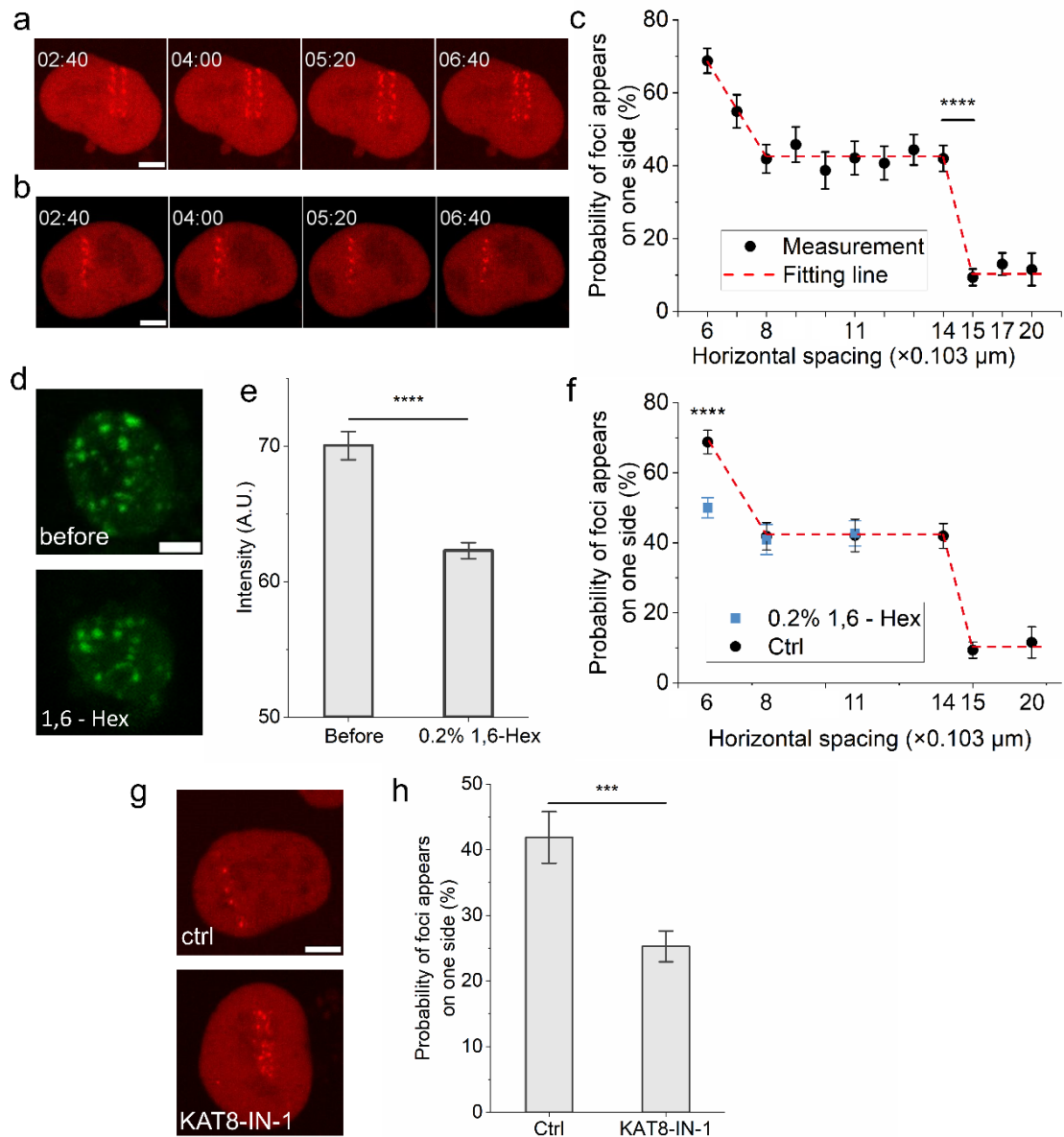

**Figure 2. Spatial and physical characteristics of DSB clustering.**
**a**, representative image of 53BP1 foci induced at each damage site by a 2×5 point laser microirradiation, with a vertical spacing of 2 μm and horizontal spacing of 2 μm; scale bar, 5 μm. **b**, Representative image showing only one single focus generated at two adjacent damage points in the 2×5 point laser microirradiation, with a vertical spacing of 2 μm and horizontal spacing of 0.6 μm; scale bar, 5 μm. **c**, Relationship between the probability of foci appears on one side and the distance, data shown as mean and s.e.m (n≥93, **** p<0.0001, independent samples t-test). **d**, Representative image of 53BP1 foci after treating cells with 4 μM etoposide for 30 minutes (top), and after 20 minutes of treatment with 0.2% 1,6-Hexanediol (bottom). **e**, Changes in foci fluorescence intensity before and after adding 1,6-Hex, data shown as mean and s.e.m (n=16). **f**, Probability of foci appears on one side under 0.2% 1,6-Hex treatment during 2×5 point laser microirradiation, data shown as mean and s.e.m (n≥132; blue: untreated cells, red: cells treated with 0.2% 1,6-Hex; **** p<0.0001, independent samples t-test). Ctrl
8

represents the measurement in Figure 2c. **g**, Representative images of the control group (top) and cells treated with KAT8-IN-1 for 30 minutes before 2×5 point laser microirradiation (bottom). **h**, The probability of foci appearing on one side is assessed for control and KAT8-IN-1-treated cells under 2×5 point laser microirradiation, with a vertical spacing of 2 μm and a horizontal spacing of 0.8 μm; data shown as mean and s.e.m (n≥345, *** $p<0.001$, independent samples t-test).

## Construction of the entire nucleus model

In our previous research, we developed a model of radiation-induced damage on a single chromatin strand[26,27,32]. To further investigate the spatio-temporal properties of DSB clustering, we constructed the entire nucleus model simulating DNA damage across the entire cell nucleus post-irradiation using Monte Carlo method. The DSB yield predicted by this model is closely aligned with prior experimental results[33] (Table 1). Additionally, the ratio of direct DNA damage and indirect DNA damage calculated by our model was approximately 4:6, which was consistent with previous reports[26,34], demonstrating the reliability of the model.

|  | SSB yield (/Gy/Gbp) | DSB yield (/Gy/Gbp) |
|---|---|---|
| Simulation | 182.3 | 5.6 |
| Botchway[33] |  | 5.6 |

**Table 1. Simulation of DSB damage induced by $^{60}$Co radiation.**

## Simulate the formation, fission and fusion of 53BP1 foci

Based on the entire nucleus model, we developed a model for the formation, fission, and fusion of 53BP1 foci. First, the model simulated DNA damage post-irradiation (Fig. 3a), where DSBs within the interaction radius of foci formation $R_{foci}$ can form a focus. At 30 minutes post-irradiation, DSB repair is initiated in the simulation. After each DSB repair event, the model reevaluates whether the remaining DSBs within the foci can maintain the original structure based on the fission interaction radius $R_{fission}$. Then, based on the fusion distance $R_{fusion}$, determine whether the nearby foci's DSBs are fused with that focus. By fitting the number of foci induced by 1 Gy $^{60}$Co irradiation, we determined the values for $R_{foci}$=950 nm (Fig. 3b). Based on the probabilities of foci fission and fusion, we adjusted and fitted $R_{fission}$=1000 nm, and $R_{fusion}$=1200 nm (Fig. 3c-e).

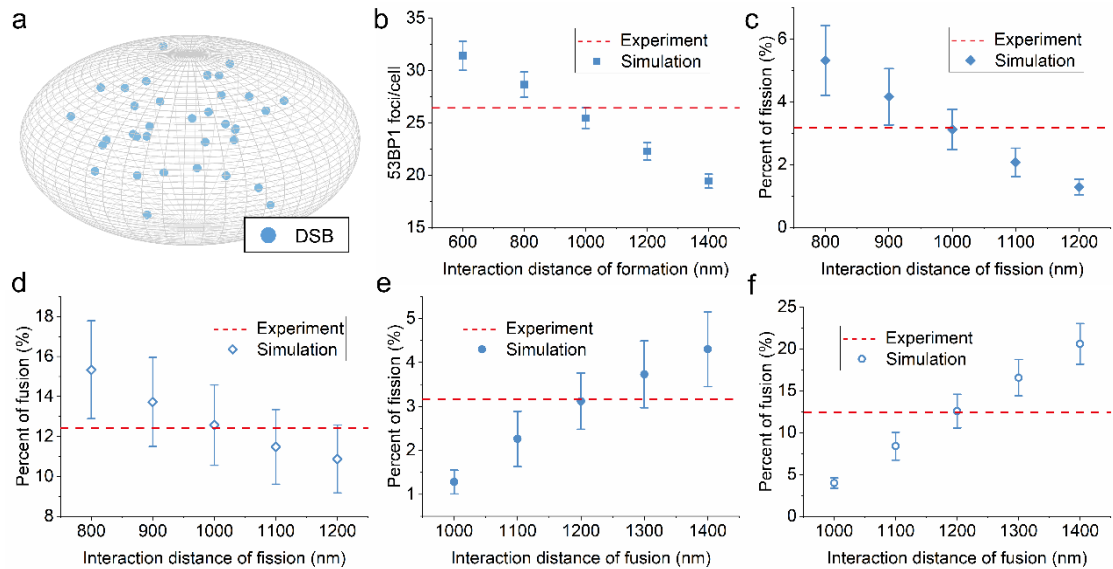

**Figure 3. Dynamics model of 53BP1 foci formation, fission, and fusion.**
**a**, Simulation of the initial distribution of double-strand breaks (DSBs) in cells exposed to 1 Gy of $^{60}$Co radiation. Blue dots represent DSBs. **b-f**, Fitting the interaction distance of 53BP1 foci formation, fission and fusion; red lines represent experimental values, while blue dots indicate fitted values; data shown as mean and s.e.m. **b**, Fitting the interaction distance of 53BP1 foci formation based on the number of foci induced by 1 Gy of $^{60}$Co radiation. **c**, Fitting the interaction distance of fission based on the fission probability of foci at 1 Gy $^{60}$Co radiation. **d**, Fitting the interaction distance of fusion based on the fusion probability. **e**, Fitting the interaction distance of fusion based on the fission probability. **f**, Fitting the interaction distance of fusion based on the fusion probability.

The above model enables the simulation of the dynamics of 53BP1 foci formation, fission, and fusion after 1 Gy $^{60}$Co irradiation. The simulation provides the probabilities of foci fission and fusion (Fig. 4a), as well as the distances between foci before fusion and after fission (Fig. 4b), which are consistent with experimental data. The simulation results showed that the percentage of foci fusion increases with foci repair (Fig. 4c). As the number of foci decreases, the frequency of fusion gradually diminishes. By assigning the DSB repair half-life as 75 minutes based on the experimental foci repair curve (Supplementary Fig. 1d), the model can successfully simulate the dynamic curve of foci numbers during DSB repair, which aligns well with experimental data (Fig. 4d). Furthermore, our model can be applied to simulate data from previous studies. By using the reported foci repair half-lives, the model can predict the dynamics of foci number[5,12,35] (Fig. 4e-g). These demonstrated its robustness in accurately simulating the dynamic processes of 53BP1 foci formation, fission, and fusion.

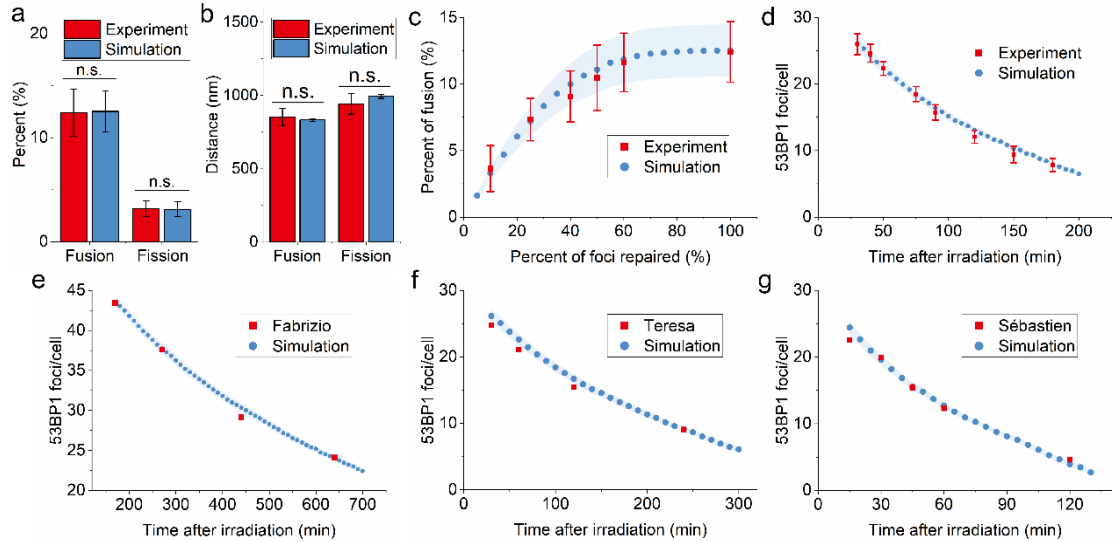

**Figure 4. Simulation of the repair dynamics of 53BP1 foci.**
**a**, Simulation of the fission and fusion probabilities of 53BP1 foci induced by 1 Gy of $^{60}$Co radiation (n.s., p>0.05, independent Samples t-test). **b**, Simulation of the distances between foci after fission and before fusion under 1 Gy conditions (n.s., p>0.05, independent samples t-test). **c**, Simulation showing the increasing proportion of foci that have undergone fusion during the repair process. **d-g**, Predition of the repair curve of 53BP1 foci; red, experimental data; blue, simulated data, data shown as mean and s.e.m. **d**, Prediction of the repair curve for 53BP1 foci in HT1080 cells in our experiment. **e**, Prediction of the repair curve for 53BP1 foci induced by 2 Gy X-rays in U2OS cells[5]. **f**, Prediction of the repair curve for 53BP1 foci induced by 1 Gy X-rays in HT1080 cells[12]. **g**, Prediction of the repair curve for 53BP1 foci induced by 1 Gy X-rays in A549 cells[35].

## Prediction of the dynamics of 53BP1 foci

To further validate the previously measured probability of DSB clustering at different distances (Fig. 2b), we incorporated this probability into the foci formation model. First, we simulated the DSB distribution in the entire nucleus model. For each DSB and its adjacent DSBs, the probability of clustering into a single focus was determined based on the distance between them, allowing us to simulate foci formation. The predicted number of foci induced by $^{60}$Co irradiation at various doses, as well as in fractionated irradiation (1 Gy + 1 Gy), closely matched the experimental results (Fig. 5a, Laser data model), with no significant differences observed in t-tests. This strongly supports the accuracy of the previously obtained data on the distance-dependent DSB clustering probability.

Based on the dynamic model, we can also predict the number of 53BP1 foci induced by $^{60}$Co irradiation at various doses (Fig. 5a), the distances between foci before fusion and after fission (Fig. 5b), and the probabilities of fission and fusion (Fig. 5c). The prediction results indicate that as the irradiation dose increases, both the number of DSBs and foci rise, and the probabilities of foci fission and fusion increase as well, while the distances of fission or fusion remain essentially unchanged. Additionally, we predicted the time curve of foci fission and fusion (Fig. 5d), as well as the foci repair curve after 2 Gy irradiation (Fig. 5e). Experimental measurements agree well with the predictions, with no significant differences in t-tests.

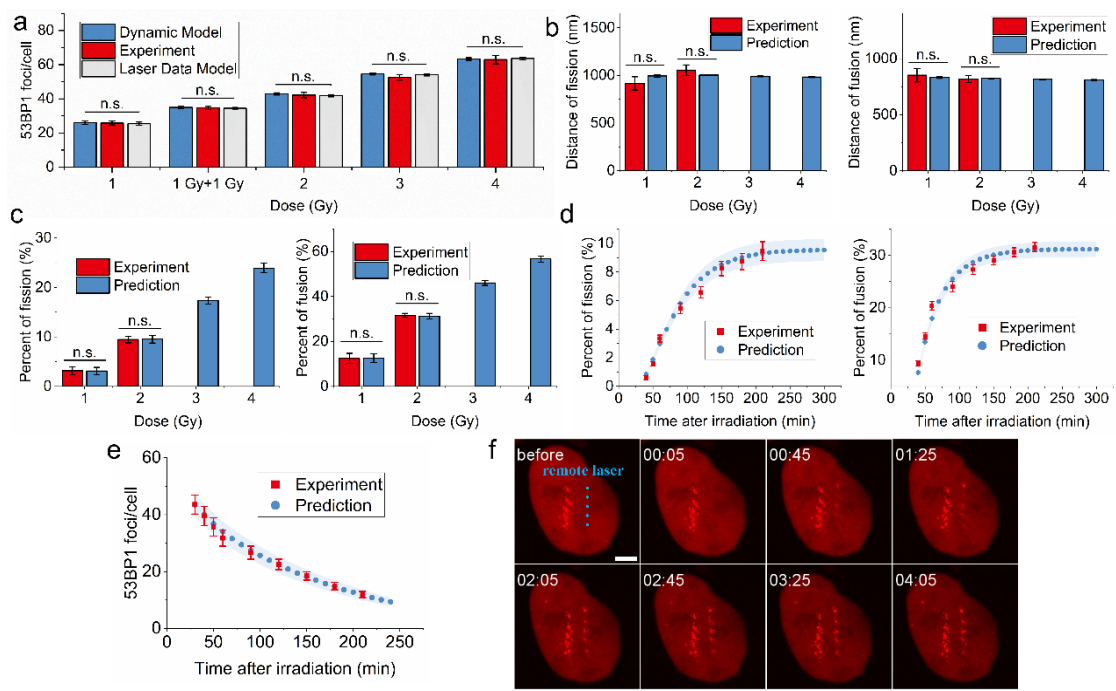

**Figure 5. Prediction of the dynamics of 53BP1 foci.**
**a**, Prediction of the number of 53BP1 foci induced by 1 to 4 Gy of $^{60}$Co radiation and fractionated irradiation; blue, the predicted values from the dynamic model; red, the experimental values; gray, the predicted values from the laser data model. **b**, Distances after fission (left) and before fusion (right) of 53BP1 foci induced by 1 to 4 Gy of $^{60}$Co radiation as predicted by the dynamic model. **c**, Prediction of the probabilities of fission (left) and fusion (right) of 53BP1 foci. **d**, Curves showing the increase in the proportion of fission (left) and fusion (right) of 53BP1 foci over time. **e**, Prediction of the repair curve of foci at a dose of 2 Gy. Blue represents predicted values, while red represents experimental values; data are shown as mean and s.e.m. **f**, Representative time series image of a column of damage points (1×5) is induced 4 μm to the right of a column of 53BP1 foci; scale bar, 5μm.

Based on the dynamic model of 53BP1 foci, we can also predict that DSBs with large distances cannot cluster. To verify this, we repeated the former experiment by first generating a vertical 1×5 dot matrix of damage sites and then generating another same 1×5 dot matrix 4 μm to the right of the first one. Theoretically, DSBs at two sides given the distance of 4 μm are unable to cluster. As predicted, a newly formed row of foci was observed on the right in the simulation (Figure 5f).

# Discussion

This study focuses on the spatiotemporal dynamics of DSB clustering. Overall, it reveals that the clustering of DSBs varies throughout the repair process and is regulated by spatial distance. Additionally, we developed a predictive model for 53BP1 foci dynamics.

Firstly, we observed that DSB clustering evolves over time, with the occurrence of primary clustering and secondary clustering separated by a plateau phase. The timing of these phases highly coincides with the stages of DSB repair. The repair rate of secondary clustered DSB was significantly slower than that of primary clustered DSB, with no significant difference compared to those non-clustered DSB (Fig. 1g). Based on these observations, we propose the following hypothesis. In the early stages of DDR, the cell nucleus hasn't prepared a full set of repair proteins for all DSBs. In situations of limited resources, DSB clustering allows multiple DSBs to share repair proteins, thereby accelerating the initiation of the repair process. This approach is more economical and efficient for the cell, aligning with the principles of biological evolution. Additionally, primary clustering may help fast establishment of a TAD-scale DDR focus and promote the formation of D compartments[10]. During the repair phase, secondary clustering of DSBs occurs. Secondary clustered DSBs tend to repair more slowly, which is consistent with the findings from François et al.[7]. Cells may recognize hard-to-repair DSBs for secondary clustering, facilitating more accurate repair. Since 53BP1 foci act as repair compartments containing a variety of repair proteins[18], secondary clustering enhances the integration of repair resources and improves repair efficiency.

We measured the probability of DSB clustering at different distances. By incorporating this data into the DNA damage model without setting any additional parameters, we were able to directly predict the number of 53BP1 foci induced by irradiation. Furthermore, we could predict the number of foci induced by fractionated irradiation. These results strongly demonstrate the relationship between DSB clustering and distance, as well as the accuracy of our model.

DSB clustering depends on distance, following a piecewise function. This suggests that clustering may be influenced by at least two factors. Within the range of 0.8–1.4 μm, the intensity of DSB clustering is nearly constant. This is potentially due to the regulation of actin filaments, since previous studies suggested the relationship between DSB clustering and actin nucleators like Arp2/3 and FMN2[7,8]. However, at distances shorter than 0.8 μm, clustering intensity increases significantly. Numerous studies have shown that 53BP1 exhibits properties of liquid-liquid phase separation[10,12,18], and it is conceivable that surface tension from 53BP1 facilitates DSB clustering. However, no direct evidence has been provided linking 53BP1 phase separation to DSB clustering. In our work, the addition of 1,6-hexanediol, which inhibits 53BP1 phase separation, and reduced DSB clustering intensity at distances of 0.6 μm, while clustering intensity at distances of 0.8–1.1 μm remained unchanged. We observed that the diameter of 53BP1 foci is approximately 0.6 μm, indicating that when DSBs are separated by less than 0.8 μm, the probability of contact between two 53BP1 foci increases substantially. The surface tension of 53BP1 promotes the fusion of two foci, thereby contributing to DSB clustering.

Our work partially resolves the previous debate about the existence of "repair centers". Evi Soutoglou et al. observed that DSBs remain spatially stable and do not cluster[36]. This may be due to the low concentration of DSB-inducing agents used in their experiments, leading to sparse DSB distribution and distances too great for clustering, which is consistent with our spatial specificity theory.

Treatment with KAT8-IN-1 can inhibit KAT8, potentially causing DNA to become more condensed and less mobile. DSBs in transcriptionally active regions are more prone to clustering, likely because the DNA in these regions undergoes frequent unwinding, becoming more relaxed and mobile.

Vadhavkar et al. developed a regional model of foci, which well fit the number of foci across different irradiation doses[28]. Building upon this, we developed a model for 53BP1 foci formation, fission, and fusion. Since our model is rigorously based on an entire nucleus DNA model, it effectively predicts the number of 53BP1 foci as well as the processes of fission, fusion, and repair at different doses. We here introduced three parameters: the interaction radius of foci formation $R_{foci}$, the fission interaction radius $R_{fission}$, and the fusion interaction distance radius $R_{fusion}$. According to the simulation, $R_{foci}$ = 950 nm, which is much larger than the average foci radius. This suggests that DSB clustering occurs in the early stage of DDR, supporting our experimental findings. $R_{fission}$ = 1000 nm, slightly larger than $R_{foci}$, indicates the difficulty for DSBs within a focus to separate, possibly because of the surface tension of 53BP1. Notably, the surface tension of 53BP1 is not particularly strong[12]. $R_{fusion}$ = 1200 nm, larger than $R_{foci}$, suggests that secondary clustering occurs during the DSB repair phase, corroborating our experimental conclusions.

# Methods

## Cell culture

The HT1080 human fibrosarcoma cell line was used, with stable transfection of 53BP1-GFP [37] and transient transfection of 53BP1-tdTomato. PtdTomato-N1-TP53BP1 (P10814) was obtained from MiaoLingBio, China.

Cells were cultured in DMEM high-glucose medium with 10% fetal bovine serum, 100 mg/ml streptomycin, 100 U/ml penicillin and 100 mg/ml hygromycin. Cells were incubated at 37 °C in an atmosphere of 95% air and 5% $CO_2$.

## Laser microirradiation-induced DNA damage

HT1080 cells grown on a confocal dish were irradiated with high spatial precision using a 365 nm pulsed nitrogen ultraviolet laser generated by the Micropoint system (Andor). This system was integrated directly with the epifluorescence pathway of the Dragonfly confocal imaging system (Andor), enabling time-lapse imaging intervals of 20 or 30 seconds, or 2.5 minutes.

## Conventional irradiation condition

Cells were irradiated by $^{60}$Co γ-ray with 9.6 x 10$^{15}$ Bq activity (Peking University).

After irradiation, the cells were placed in a live cell workstation maintained at 37°C with 95% air and 5% CO$_2$. For cells that required fixation, 4% paraformaldehyde was used. Images were captured using an LSM 700 confocal microscope (ZEISS), with z-stacks collected at intervals of 600 nm.

## KAT8 inhibition

For KAT8 inhibition, cells were pre-treated with 5 μm KAT8-IN-1 for at least 30 minutes prior laser microirradiation. KAT8-IN-1(HY-W015239) was obtained from MedChemExpress (Monmouth Junction, NJ, USA).

## Imaging data processing

Fluorescent images were processed using ImageJ or Imaris, and the count, size, fluorescence intensity, and position of 53BP1 foci were quantified.

## Construction of 46 chromosomes model of cell nucleus model

Chromosomes have a hierarchical structure. Starting from the double helix structure, we constructed nucleosomes, 30-nm fibers, chromatin territory and other structures in sequence, and finally constructed each chromosome. Then determine the position of each chromosome in the nucleus according to the gene density of each chromosome.

According to geometric characteristics such as the pitch and rotation of the double helix structure, the double helix structure can be divided into A type, B type, Z type, etc. In our model, the most common structure, B-DNA double helix, was used[26]. This structure was with 2.0 nm diameter and 0.34 nm pitch. We set 10 ten nucleotide pairs per helical turn to be consistent with the experimental value of 10–10.5 pairs per turn[38]. We used the small sphere to represent the base and phosphoribose.

The basic unit of the higher-order structure of eukaryotic chromatin is the nucleosome, which consists of 146 pairs of nucleotides wrapped around a histone octamer about 1.7 times, plus linker DNA[39]. In our model, since the histone octamer does not directly interact with the radiation, it could be simplified as a sphere with a diameter of 6.5 nm. The winding diameter was set to 9 nm, so that the diameter of the nucleosome was 11 nm, which was consistent with experiments[38]. The pitch of each winding was set to 2.6 nm.

30-nm chromatin fibers were used to fill in nucleosome-level chromatin structures. Nucleosomes in the 30nm chromatin fiber were as zig-zag structure[40]. The basic unit of chromatin was clutch[41], a group of 12 nucleosomes (Supplementary Fig. 3a). Then we used the chromosome territories and interchromatin compartment (CT–IC) model to construct interphase chromosome[42]. In our model, the interchromatin compartment was represented as a sphere 500 in diameter containing 1 Mbp (Supplementary Fig. 3b). According to research on human cells, the radial distribution of chromosomal territories within the cell nucleus is related to the DNA content and chromosome size of the chromosome, with other influencing factors, such as transcriptional activity, replication time, GC content[43]. We assumed that DNA content and chromosome size both affected the radial distribution of chromosomal territories[44]. Consist with Kerth, in our model, the distribution of chromosome i in the nucleus obeyed the following formula[44],

$$P(d)_i = \exp\left(\frac{density(No.i)}{density(No.19)}\right) \cdot \alpha$$

$P(d)_i$ represented the probability that the position of chromosome i was d, which was the distance from chromosome to nucleus center. $density(No.i)$ represents the gene density of chromosome i. Chromosome 19 is the chromosome with the highest gene density among human 46 chromosomes. $P(No.19)_i$ represented probability that the position of chromosome 19 is d. α is the parameter of the adjustment program to ensure that the distribution of chromosomes in the nucleus is not too dense or too loose. Under this distribution law, the chromosomes with higher gene density are closer to the center of the cell nucleus.

Then, we randomly arranged chromosomes with their center obeying above formula. The initial state of the chromosomes was a spirally ascending rod-shaped spherical chromatin territory (Supplementary Fig. 3c). Each circle has 6 spherical chromatin compartments with a pitch of 252 nm. The length of each chromosome generated was from 1.9 to 10.5 μm.

The distance between the compartments is mainly affected by the three potential energy $U_S$, $U_E$, $U_B$[44]:

$$U_S(r) = \frac{3K_BT}{l_0^2}r^2$$

$$U_E(r) = U_0\left(1 + \frac{r^4 - 2D^2r^2}{D^4}\right)$$

$$U_B(r) = \begin{cases} 0, & r < R_{Terr} - \frac{D}{2} \\ \frac{U_0}{5D}\left(r - R_{Terr} - \frac{D}{2}\right), & r \geq R_{Terr} - \frac{D}{2} \end{cases}$$

However, relying on these three potential energies can only reflect the spatial influence between compartments, without considering the limiting effect of the karyotheca on chromatin. So, we introduced forth potential energy, karyotheca restriction,

$$U_K = U_{K_0}(1 - t^2)$$

$$t = \sqrt{\left(\frac{x}{a - \frac{D}{2}}\right)^2 + \left(\frac{y}{b - \frac{D}{2}}\right)^2 + \left(\frac{z}{c - \frac{D}{2}}\right)^2}$$

x, y, z is spatial coordinates of the chromatin compartment. a, b, c is spatial coordinates of semi-axis length of ellipsoid (cell nucleus). For $t > 1$, it meant that the chromatin compartment was partly or fully out of the cell nucleus. At this time, it would be restricted by a traction force within the cell nucleus.

Then, we let each chromatin compartment randomly walk until equilibrium. Each time move one chromatin compartment, the direction and step length were random. And each movement cannot exceed 500 nm. Calculate the overall energy change after the random walk. If the overall energy decreases after the movement, accept this step. If the overall energy rose after the movement, accept this step with the probability of $\exp(-\Delta E)$, otherwise rejected. We set $U_{K_0}$ to $1 \times 10^7\ kT$, and T to 310 K. Repeated the above steps 600000 times to make the system reach a balanced state (Supplementary Fig. 3d). Finally, most chromatin compartments were inside the nucleus (Supplementary Fig. 3e).

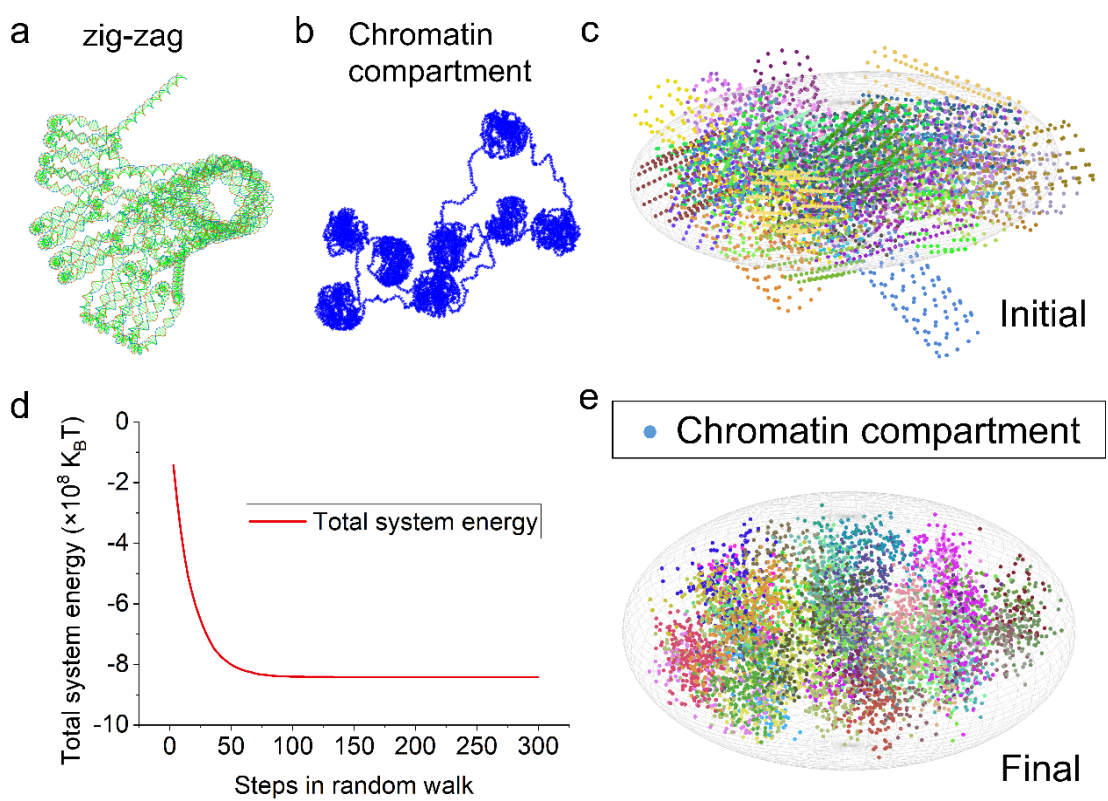

**Supplementary Figure 3. Construction of the entire nucleus model**
**a**, The zig-zag structure is used as the fundamental unit of chromatin. **b**, Chromatin compartments. **c**, Initial arrangement of chromatin compartments; one color represents one chromatin strand. **d**, The total energy of the system varies with the number of steps in the random walk. **e**, After the random walk, the final arrangement of chromatin compartments.

## Simulation of ionizing radiation

We used Geant4 toolkit with Geant4-DNA processes to simulate track structure of ionizing radiation[45,46]. Simulation procedure and detailed parameter setup were consisted with previous report[26,47].

## Calculation of DNA damage

We superimposed the chromatin structure with simulated track structure to calculate DNA damage. Detailed procedure and principle were reported previously[48].

## Model for the formation, fission and fusion of 53BP1 foci

Using the entire nucleus model, the spatial distribution of DSBs generated by irradiation can be simulated. The effective distance $R_{foci}$ for foci formation is set, where DSBs within this distance will form a focus early in the DDR[28]. After passing through a stationary phase, DSBs begin to be repaired. The effective distance for fission, $R_{fission}$, and for fusion, $R_{fusion}$, are established. If a DSB exceeds the effective distance for fission, the foci will undergo fission. If DSBs from different foci are within the effective distance for fusion, they will merge into a single focus.

## Acknowledgements

We thank the Core Facilities at the School of Life Sciences, Peking University, for assistance with the Dragonfly.